# Quantitative Susceptibility Mapping through Model-based Deep Image Prior (MoDIP)

Zhuang Xiong[1], Yang Gao[2, 1], Yin Liu[3], Amir Fazlollahi[4], Peter Nestor[4], Feng Liu[1], Hongfu Sun[1]*

1. School of Elecrical Engineering and Computer Science, University of Queensland, Brisbane, Australia
2. School of Computer Science and Engineering, Central South University, Changsha, China
3. The Third Xiangya Hospital, Central South University, Changsha, China
4. Queensland Bain Institute, University of Queensland, Brisbane, Australia

*Abstract*—The data-driven approach of supervised learning methods has limited applicability in solving dipole inversion in Quantitative Susceptibility Mapping (QSM) with varying scan parameters across different objects. To address this generalization issue in supervised QSM methods, we propose a novel training-free model-based unsupervised method called MoDIP (Model-based Deep Image Prior). MoDIP comprises a small, untrained network and a Data Fidelity Optimization (DFO) module. The network converges to an interim state, acting as an implicit prior for image regularization, while the optimization process enforces the physical model of QSM dipole inversion. Experimental results demonstrate MoDIP's excellent generalizability in solving QSM dipole inversion across different scan parameters. It exhibits robustness against pathological brain QSM, achieving over 32% accuracy improvement than supervised deep learning and traditional iterative methods. It is also 33% more computationally efficient and runs 4 times faster than conventional DIP-based approaches, enabling 3D high-resolution image reconstruction in under 4.5 minutes.

## I. INTRODUCTION

MAGNETIC susceptibility is a tissue property that can be used to quantify changes in myelin [1], iron [2], and calcium [3] concentrations. It serves as a potential biomarker for studying neurological diseases, including Alzheimer's Disease [4], Parkinson's Disease [5], Huntington's Disease [6], Multiple Sclerosis [7], [8] and cerebral vascular injuries such as microbleeds [9] and hemorrhages [10], [11], [12]. Additionally, it has been applied in brain function studies by quantifying brain oxygen level variations [13], [14], [15], [16], [17].

To reconstruct the susceptibility map from a Magnetic Resonance Imaging (MRI) phase measurement, a technique known as Quantitative Susceptibility Mapping (QSM) [18], [19] [20], involving multiple image processing steps, has been developed. These steps include phase unwrapping, background field removal, and dipole inversion. The final dipole inversion step computes the susceptibility map by performing a deconvolution operation with the unit dipole kernel, which raises the challenge of an ill-posed inverse problem. Several iterative optimization approaches have been developed to solve this problem [21], [22], [23], [24], which incorporate image regularization techniques to suppress noise and streaking artifacts. For instance, the MEDI method [25] assumes the predicted QSM shares tissue anatomical boundaries with the local field map. The SFCR method [26] refines the morphological prior by including susceptibility structural features. The iLSQR method [27] solves QSM dipole inversion by estimating and removing streaking artifacts from initial LSQR estimation. STAR-QSM [28] further improves susceptibility inversion for image data with high-intensity sources. There are also total field inversion [29], [30] and single-step-QSM [31] methods, which directly derive susceptibility maps from the unwrapped or raw phase maps. However, these conventional regularization techniques often result in artefactual, blurry, and underestimated susceptibility maps. Additionally, they often require fine-tuning of the regularization parameters to adapt to varying scan parameters and objects.

Deep learning methods have been developed to solve dipole inversion in QSM. However, most supervised learning methods [32], [33], [34], [35] have been trained solely on 1 mm isotropic resolution image and pure axial acquisition datasets of the human brain. As a result, their applicability is restricted when dealing with varying spatial resolutions, acquisition orientations, brains with abnormal sources (e.g., high-intensity hemorrhages), or objects other than the human brain. To enhance model generalizability, some methods [36], [37], [38] have employed synthetic data and meta-learning strategies, while other methods [39], [40], [41] have incorporated deep neural networks in an iterative manner guided by the physical model of dipole inversion. A recent study [42] has also embedded affine transformations into an end-to-end model. While these supervised methods have demonstrated improvements in model generalization, their performance remains constrained by the availability and diversity of training datasets. In contrast, a self-supervised solution [43] has been proposed, which exhibits greater robustness against image resolution variations through adaptive instance normalization; however, the acquisition orientation effect was not investigated.

Deep Image Prior (DIP) [44] introduced deep neural networks as an implicit regularization technique for solving inverse problems. In the context of QSM, a recent approach called FINE (Fidelity Imposed Network Edit) [45] demonstrated the capabilities of DIP by fine-tuning pre-trained networks during the inference stage. However, this study highlighted the intensive forward and backward computations through the large U-net architecture, consuming substantial GPU memory and computational time. As a result, FINE imposed significant limitations on image size and resolution compared to conventional supervised methods.

In contrast to conventional DIP-based methods like the original DIP [44] and FINE [45], which heavily rely on protracted iterations and intricate network architectures for image reconstruction, our proposed method, Model-based DIP (MoDIP), takes a different approach. It combines a compact mini-U-net with a Data Fidelity Optimization (DFO) module, enhancing generalization capabilities while substantially reducing computational complexity. The main contributions can be summarized as follows:

- We introduced MoDIP, a novel deep learning-based method that combines physical model guidance with a mini-U-net architecture in an unsupervised manner. Unlike existing physics-based unrolled methods, MoDIP does not require any training data.
- We demonstrated through theoretical analysis and experimental results that the synergy of the mini-U-net and the DFO module leads to well-regularized image reconstruction with faster convergence, lower memory requirements, and mitigated overfitting.
- We validated MoDIP's improved generalizability and superior reconstruction quality and accuracy by applying it and comparing it with other deep learning and conventional methods on various test datasets of different imaging objects, value ranges, and measurement conditions.

## II. RELATED WORK

### A. QSM Dipole Inversion

The perturbated local field $\varphi$ due to tissue susceptibility $\chi$ can be formulated as a convolution with the unit dipole kernel $D$. This physical model can be written in k-space using the following equation:

$$\varphi = \mathcal{F}^{-1}D\mathcal{F}\chi, \quad (1)$$

where $\mathcal{F}$ and $\mathcal{F}^{-1}$ denote the forward and inverse Fourier transforms. The unit dipole kernel $D$ can be represented as:

$$D(\vec{k}) = \frac{1}{3} - \frac{(p_x k_x + p_y k_y + p_z k_z)^2}{k_x^2 + k_y^2 + k_z^2}, \quad (2)$$

Here $k = [k_x, k_y, k_z]$ denotes the k-space coordinates and $\vec{p} = [p_x, p_y, p_z]$ is the vector projections of the field-of-view onto the main magnetic field vector $\vec{B_0}$ [46]. The relationship between the image voxel size ($\vec{v} = [v_x, v_y, v_z]$) and the k-space coordinates is given by:

$$k_x = \frac{x}{M_x v_x}, \; k_y = \frac{y}{M_y v_y}, \; k_z = \frac{z}{M_z v_z} \quad (3)$$

where $[M_x, M_y, M_z]$ denotes the image matrix size.

A minimization task can be formulated to solve the susceptibility $\chi$ from the local field $\varphi$:

$$\chi^* = \arg\min_{\chi} \left[ \|\mathcal{F}^{-1}D\mathcal{F}\chi - \varphi\|_n + \lambda * \mathcal{R}(\chi) \right], \quad (4)$$

where $\mathcal{R}(\chi)$ is a regularization term to penalize the data fidelity, and $\lambda$ is the weighting factor for the regularization. Conventionally, the regularization term often originates from prior assumptions such as sparsity and smoothness. However, such assumptions often yield suboptimal results and require manual parameter tuning for different data.

### B. Supervised Learning for QSM

Supervised deep learning dipole inversion approaches try to learn the optimal parameter $\theta$ under a pre-defined model $f$, for mapping from local field to tissue susceptibility:

$$\theta^* = \arg\min_{\theta} \sum_{i=1}^{N} \|f_\theta(\varphi_i) - \chi_i\|_n, \quad (5)$$

where $\varphi_i$ and $\chi_i$ are pairs of inputs (local field maps) and labels (tissue susceptibility maps) for training. Data-driven intrinsic regularization is learned by fitting the entire training dataset. However, most existing supervised QSM methods are trained using 1 mm isotropic resolution images acquired in a pure-axial head orientation ($\vec{p} = [0, 0, 1]$). Consequently, when applying these trained models to testing data with scan parameters different from the training dataset (e.g., anisotropic resolutions or oblique acquisitions [42]) or in out-of-distribution cases such as lesions with substantially higher susceptibility values than healthy brain tissue [28], [35], [47], the networks may fail to yield accurate results.

### C. Deep Image Prior (DIP)

Deep Image Prior (DIP) [44] proposed to optimize the network weights $\theta$ by "fitting" a single inference data point without the need for any pre-training. For QSM dipole inversion, the DIP optimization can be written as:

$$\theta^* = \arg\min_{\theta} \|\mathcal{F}^{-1}D\mathcal{F}f_\theta(z) - \varphi\|_n, \quad (6)$$

where $z$ is the network input (Gaussian noise as in the original DIP paper). The network architecture implicitly provides regularization by acknowledging that noise in the measurement is more challenging for the network to fit than the underlying clean data. However, the untrained nature of the network hinders the application of DIP in terms of reconstruction time, which typically requires hundreds to thousands of iterations. Moreover, it is susceptible to being trapped in local minimums when sub-optimal hyper-parameters are set, as discussed in the original paper [44].

### D. Fidelity Imposed Network Edit (FINE)

An alternative approach, FINE (Fidelity Imposed Network Edit) [45], has been proposed. This method involves training a deep neural network using paired local field and QSM datasets and fine-tuning the network weights for each local field input during inference. While sharing similarities with DIP, FINE requires supervised pre-training. In essence, FINE can be seen as a particular case of DIP, where the network is initialized with pre-trained weights instead of random initializations. When the pre-training and testing datasets closely align, the pre-trained FINE model is expected to perform more effectively than DIP's untrained model starting from scratch. However, FINE's performance diminishes when the pre-training datasets deviate from the refinement scenario, as we will show in our study. Moreover, the computational demand of the large network used in FINE limits its application for higher-resolution 3D volumes.

## III. THEORY AND METHODS

### A. Model-based Deep Imaging Prior (MoDIP)

We propose a Model-based Deep Image Prior (MoDIP) method for improving the reconstruction speed, memory efficiency, and accuracy of the original DIP approach in QSM. As depicted schematically in Fig. 1, the bottom grey plane represents the manifold of QSM solutions to the local field, which satisfies the exact model loss of Eq. 1. Due to measurement noise and errors in the local field map, the desired optimal solution $\chi^*$ results in a non-zero cost and is, therefore, off this plane, as indicated by the red dot. Pure optimization with no prior (brown curve) and DIP with deep and large network architectures (dark blue curve) may converge to solutions in the zero-cost plane, far away from $\chi^*$, resulting in an overfitting effect that amplifies noise and errors in the reconstructed QSM images.

The proposed MoDIP method combines the concept of DIP with physics model-based optimization. In MoDIP (green solid curve), the network adjusts its trajectory and converges to an initial estimation $\chi_0$ of QSM. This network output serves as the starting point for the subsequent Data Fidelity Optimization (DFO) process (green dash curve), leading to the final QSM prediction $\chi_n$. The DFO process minimizes the QSM physical model objective, bringing the solution closer to its optimum. In MoDIP, the network's task is to transform the local field into an easily attainable interim state instead of directly converting it into QSM. This approach relieves the network from the excessive burden of dipole inversion and accelerates its convergence, enabling the use of a light-weight network to reduce computational costs and GPU memory requirements, thus facilitating the reconstruction of high-resolution 3D images. It is worth noting that, unlike the unrolled methods [39] [40], MoDIP performs only a single forward pass through the network, followed by multiple rapid and memory-efficient gradient descent optimization steps within each iteration.

Furthermore, we replace the Gaussian noise input in the original DIP paper with the local field map, which shares some image features with the susceptibility map, easing the network's transformation task. Lastly, in addition to the physical model loss, we introduce a local field Laplacian loss to promote tissue boundary consistency.

---

**Algorithm 1:** Model-based Deep Image Prior (MoDIP)
**Require:** $\varphi$
1: **repeat**
                ▷ network output
2:  $\chi_0 = f_\theta(\varphi)$
            ▷ model-based optimization
3:  **for** $i = 0$ **to** $n - 1$ **do**
4:    $\chi_{i+1} \leftarrow \chi_i - \alpha \cdot \nabla_{\chi_i} \|\mathcal{F}^{-1}D\mathcal{F}\chi_i - \varphi\|_2^2$
5:  **end for**
6:  **return** $\chi_n$
             ▷ network weights update
7:  **update** $\theta$ on
    $\nabla_\theta \left[ \|\mathcal{F}^{-1}D\mathcal{F}\chi_n - \varphi\|_1 + \|\nabla^2\mathcal{F}^{-1}D\mathcal{F}\chi_n - \nabla^2\varphi\|_1 \right]$
8: **until** converged

---

The overall methodology of MoDIP is illustrated in Fig. 2, which depicts the DIP process of updating the network parameters $\theta$ followed by the optimization process of updating the susceptibility map $\chi$. Detailed implementations are described in Alg. 1. Firstly, an initial estimation of the susceptibility map, denoted as $\chi_0$, is predicted from the local field $\varphi$ using a mini-U-net $f_\theta$:

$$\chi_0 = f_\theta(\varphi). \qquad (7)$$

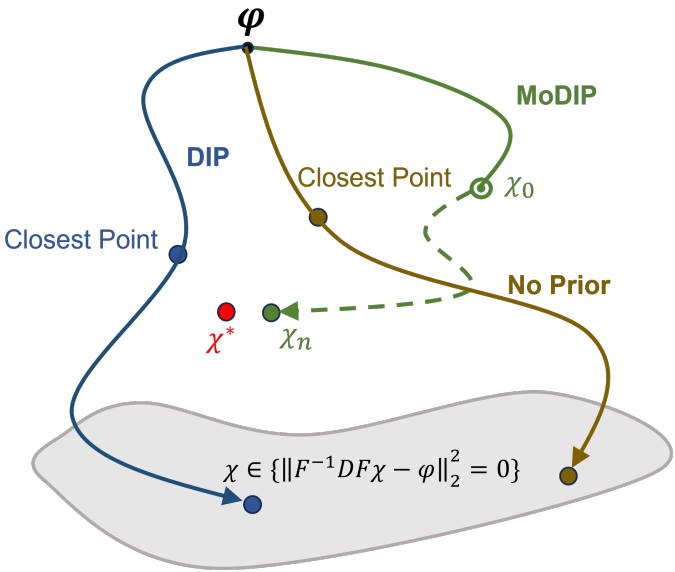

Figure. 1. Visual representation of QSM reconstruction using different approaches: DIP with heavy-weight (dark blue) networks, conventional image optimization (brown curve), and the proposed MoDIP method (green curve). $\chi^*$ signifies the optimal QSM solution, while $\chi_0$ and $\chi_n$ correspond to MoDIP's network output and its subsequent optimization result, respectively. The bottom grey plane represents the manifold where solutions satisfy the formulation.

Following the forward pass through the network, $n$ steps of gradient descent optimizations are performed on the data fidelity term and the final prediction $\chi_n$ is computed:

$$\chi_{i+1} = \chi_i - \alpha \cdot \nabla_{\chi_i} \|\mathcal{F}^{-1}D\mathcal{F}\chi_i - \varphi\|_2^2, \qquad i = 0 : n-1, \quad (8)$$

On the contrary, DIP and FINE employed a heavy-weight U-net architecture with a pooling depth of 4. In the case of DIP, the initial network weights are assigned randomly. In contrast, FINE initializes the U-net weights with those pre-trained using 1 mm isotropic QSM and local field pairs in pure-axial head orientation. The number of learnable parameters and computational costs of MoDIP, DIP and FINE methods are where $\nabla$ denotes the gradient operation, and $\alpha$ is the step size for the gradient descent algorithm. Subsequently, a backward pass is performed to update the network parameters $\theta$ by minimizing the loss function:

$$\mathcal{L} = \|\mathcal{F}^{-1}D\mathcal{F}\chi_n - \varphi\|_1 + \|\nabla^2 \mathcal{F}^{-1}D\mathcal{F}\chi_n - \nabla^2 \varphi\|_1 \quad (9)$$

where $\nabla^2$ denotes the Laplacian operation. The first term represents the Mean Absolute Error (MAE) loss of the physical model, and the second term denotes the MAE loss of the Laplacians calculated for the predicted and measured local field maps.

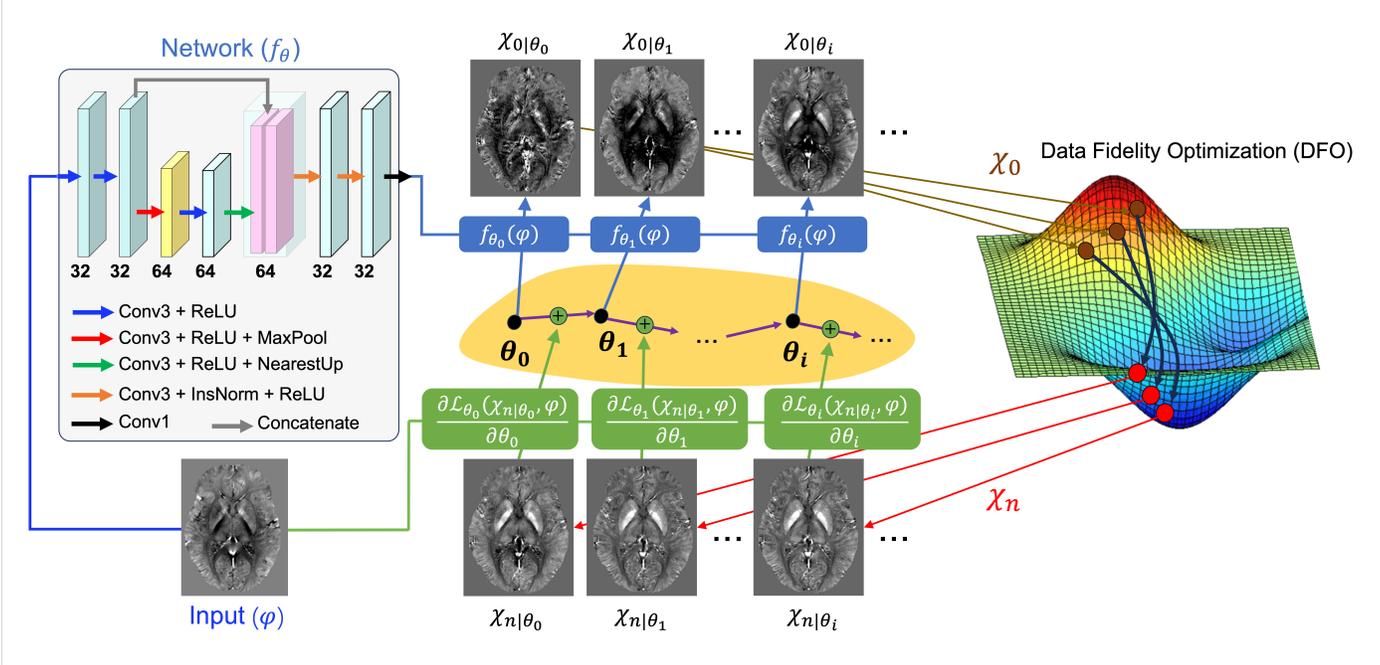

Figure. 2: The overall scheme of MoDIP for QSM reconstruction. The yellow plane represents the parameter space of the network ($\theta$), while the 3D surface illustrates the physics model cost function. An initial QSM estimate ($\chi_0$) is produced by the network and optimized by the physics model to obtain the predicted QSM ($\chi_n$). The network weights are updated based on the loss ($\mathcal{L}$) calculated on $\chi_n$. The top left of the figure shows the mini-U-net architecture, with blocks representing intermediate feature maps, and arrows indicating different operations. The number of feature maps is specified at the bottom of each block.

### B. Network Design and Implementation

MoDIP in this study utilized a mini-U-net architecture and was initialized with untrained random weights. As shown in Fig. 2, a simple encoder-decoder architecture was designed with a skip connection, similar to the modified U-net in [35], but with a pooling depth of 1 and channel numbers starting from 32. This mini-U-net consists of 8 convolutional blocks and 1 concatenation operation. For all convolutional operations, a kernel size of 3×3×3, a stride of 1, and a zero-padding of 1 are defined, except for the last one, which uses a kernel size of 1×1×1 to reduce feature dimension back to 1.

reported in Fig. 3. The evaluations are performed on two sizes of testing data: the original matrix size of 256×256×128 and the reduced size of 144×144×128. It is evident from the bar chart that MoDIP has 98% fewer network parameters, requires 33% less memory, and runs 28% faster compared to the DIP and FINE methods. The reconstruction time reported in Fig. 3 corresponds to running each method for 200 iterations.

The Adam optimizer was chosen for the optimizations of DIP, MoDIP, and FINE, with an initial learning rate of $5 \times 10^{-4}$ and a decay factor of 0.8 every 50 iterations. To ensure reproducibility, we manually fixed the random seed for model initialization. All experiments were conducted on a computer with an Intel 12700KF CPU, 32GB RAM, and an RTX4090 GPU with 24 GB vRAM. After thorough tests on different combinations of hyperparameters, we empirically set the step size ($\alpha$) to 1.2 and the number of gradient descent steps ($n$) to 10, which yielded the most favourable results, and still preserve the computational efficiency.

## IV. EXPERIMENTS AND RESULTS

### A. Ablation and Comparison Studies

To investigate the effect of the added DFO module in MoDIP, we conducted an ablation study on an in-vivo brain dataset acquired at 3T with a multi-echo GRE sequence of 1 mm suggests that the mini-U-net in MoDIP transforms the local field into an interim state instead of directly attempting the significantly more challenging task of transforming it into QSM. This is consistent with the theory depicted in Fig. 1, where MoDIP's interim $\chi_0$ output serves as an implicitly regularized starting point for the subsequent DFO process. DIP alone is unable to reach the optimal QSM solution directly from the local field in 200 iterations, while performing DFO steps alone from the measurement for extensive iterations with no prior regularization (i.e., Pure-DFO) can lead to noise and error amplification, as shown in Fig. 4 (bottom row).

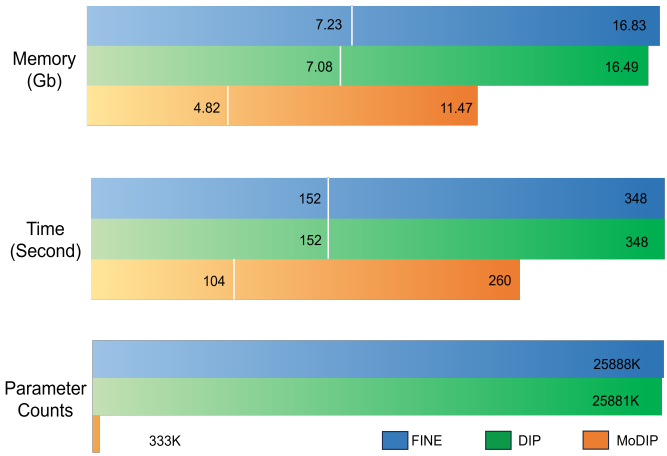

Figure. 3. Comparisons of parameter counts, memory, and time costs for FINE, DIP, and MoDIP. Black numbers indicate resource requirements. Initial segments represent results from an image size of 144x144x128, and complete bars portray results from an image size of 256x256x128.

In Fig. 5, we conducted a comparison between MoDIP, DIP, and FINE. Different network inputs (Gaussian noise vs. local field) for DIP and MoDIP were also compared. These three methods were evaluated on a simulated digital brain phantom. A COSMOS map, reconstructed from 5 GRE acquisitions with different head positions (as detailed in [29]), was resampled to isotropic resolution in pure axial orientation and processed using the standard QSM pipeline, including phase unwrapping with the best-path method [48] and background field removal using the RESHARP method [49]. We showed the intermediate results of MoDIP before DFO, referred to as $\chi_0$, and the result after DFO as $\chi_n$. These results were also compared with DIP and Pure-DFO (i.e., conventional optimization on data fidelity loss with no prior regularization). As observed in Fig. 4, unlike DIP, which targets QSM solutions, $\chi_0$ in MoDIP converges to an image that does not represent a typical QSM contrast but retains some characteristics of the local field input. This an anisotropic resolution of 1×1×2 mm³ to serve as the susceptibility ground truth. This map was used to generate a local field map, with a titled acquisition angle ($\vec{p} = [0.5, 0.5, 0.71]$), as shown in Fig. 5(a) according to Eq. 1. Each method was conducted with 500 iterations, and their model losses (Eq. 9) were plotted in Fig. 5(b). Reconstructed QSM images at iteration numbers 10, 20, 50, 100, and 200 are shown in Fig. 5(c), with error maps of the results from 200 iterations in the last column and NRMSE values at the bottom.

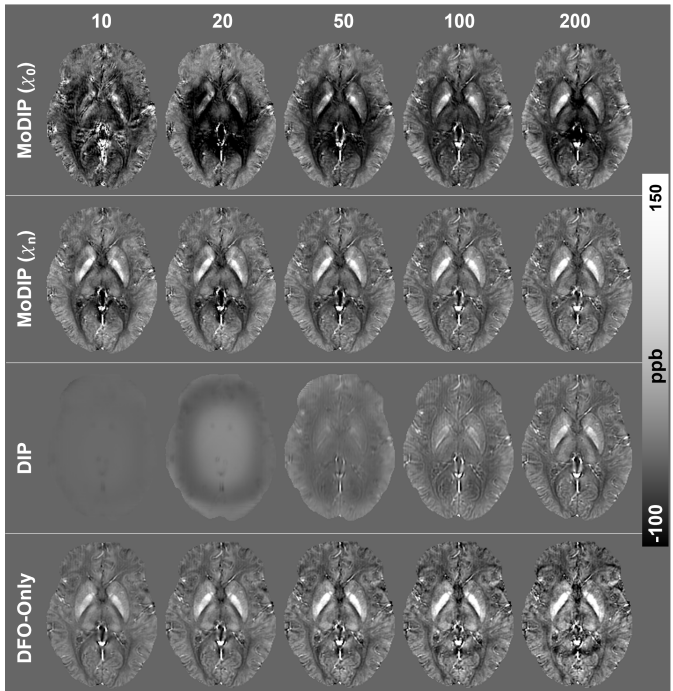

Figure. 4. Ablation study conducted on an in-vivo human brain. MoDIP intermediate results (MoDIP ($\chi_0$)) and the final outputs (MoDIP ($\chi_n$)) were visually compared to DIP and DFO-Only for 10, 20, 50, 100 and 200 iterations. The local field map was used as the input for all different methods.

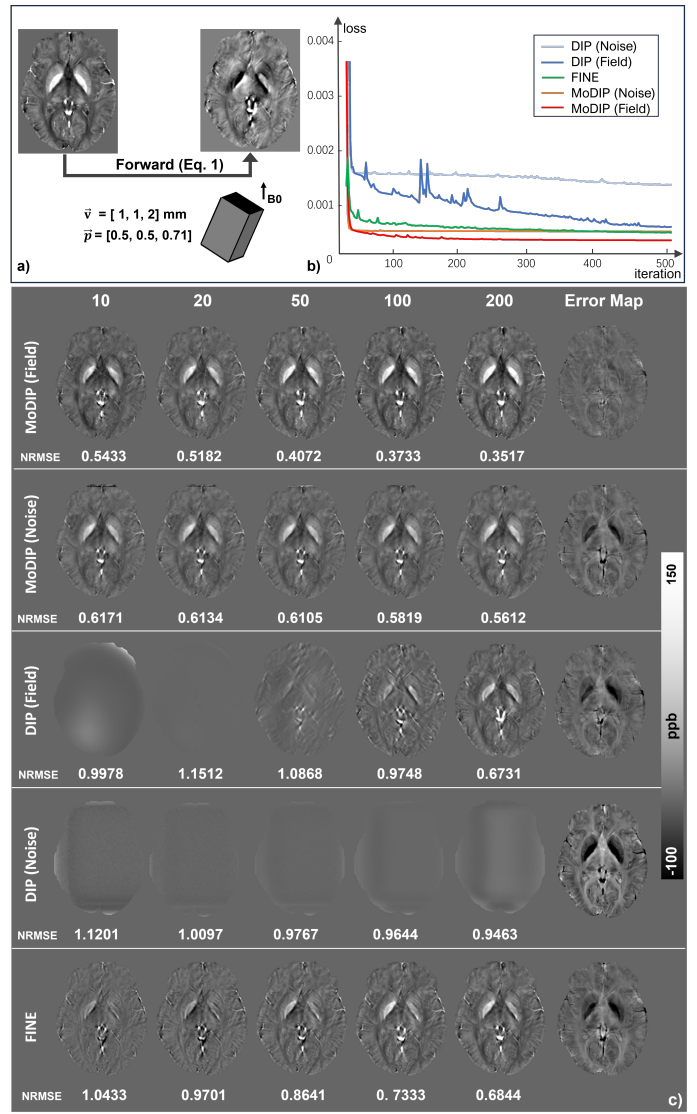

Figure. 5. a) Oblique local field simulation procedure with the given voxel size $\vec{v} = [1, 1, 2]$ mm and acquisition orientation $\vec{p} = [0.5, 0.5, 0.71]$. b), Iteration loss curves for MoDIP, DIP and FINE methods with different network inputs. c) QSM images from iterations 10, 20, 50, 100 and 200. Error maps computed between the label and the 200th iteration results for each method are shown in the final column.

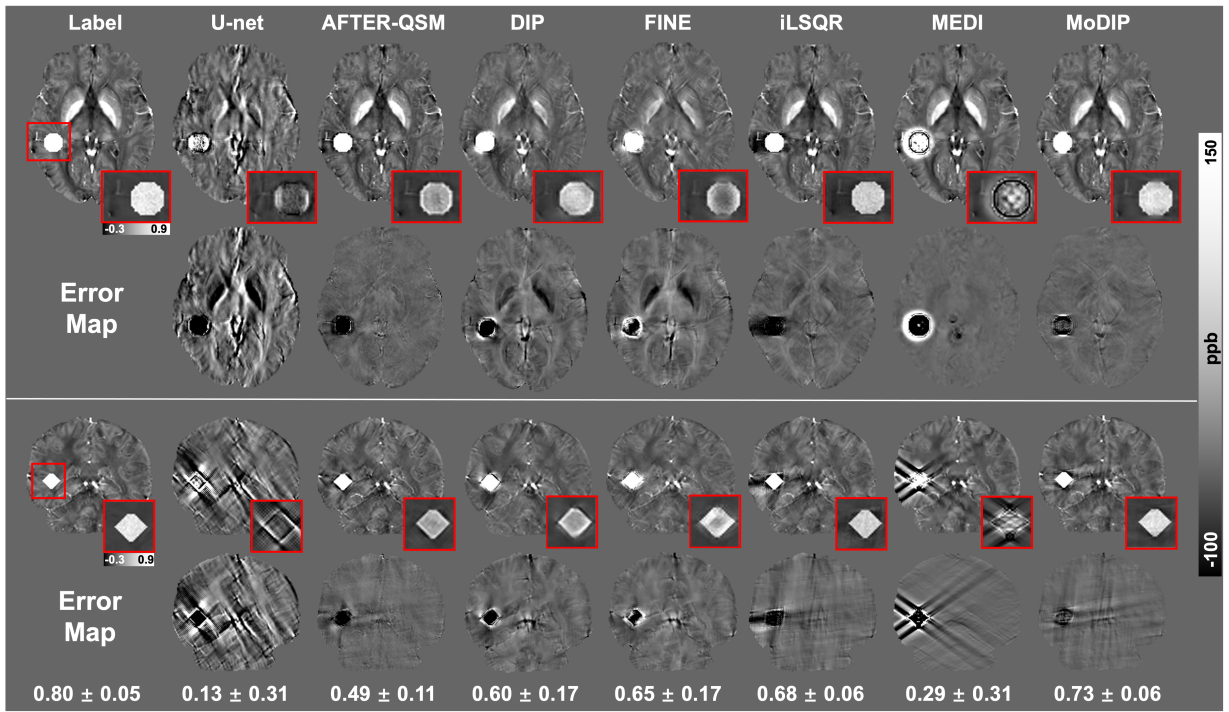

Figure 6. Comparison of various methods on a simulated brain ($\vec{v} = [1, 1, 2]$ mm, $\vec{p} = [0.5, 0.5, 0.7071]$) depicting a spherical hemorrhage source (highlighted within a red box). Results and error maps are shown in both axial and coronal views. Hemorrhage areas are magnified with window levels adjusted for best visualization. The means and standard deviations of hemorrhage susceptibility measurements are reported at the bottom.

The results in Fig. 5 demonstrate that MoDIP exhibits significantly faster convergence and achieves a substantially smaller NRMSE than DIP and FINE. MoDIP already produces a reasonable estimation of QSM with just 10 iterations, while DIP and FINE require at least 200 iterations to achieve similar performance, demonstrating the effectiveness of the added DFO module. MoDIP with Gaussian noise as the network input showed a dramatically downgraded performance compared to using the local field as the input. DIP with noise input failed to reconstruct reasonable QSM results. These suggest that the network's transfo rmation task benefits from the shared image features between the local field and the susceptibility map. For the rest of the paper, DIP and MoDIP results are all from the local field as input. Even though the local field in the testing dataset differs in image resolution and acquisition orientation from the training dataset, the pre-trained FINE method exhibited a clear benefit of faster convergence compared to the untrained DIP method.

### B. Simulated and In-vivo Pathological Brains

We compared different methods on an anisotropic ($\vec{v} = [1, 1, 2]$ mm) and tilted ($\vec{p} = [0.5, 0.5, 0.71]$) digital human brain phantom containing a simulated spherical hemorrhagic lesion with a radius of 2 mm and susceptibility of $0.8 \pm 0.05$ ppm. Results in Fig. 6 demonstrate that MoDIP outperformed all other methods visually, followed by AFTER-QSM, exhibiting excellent tissue contrast for deep grey matter, white matter, and hemorrhage, with minimal artifacts across the entire brain. Conversely, the supervised U-net failed to achieve the task, while DIP and FINE substantially suppressed QSM contrast. Susceptibility measurements of the hemorrhage are reported at the bottom of Fig. 6, with MoDIP being the closest to the ground truth. Quantitative evaluation in Table I further confirms that MoDIP achieved the minimum deviations from the ground truth in the hemorrhage region (indicated by the red box in Fig. 6), the non-hemorrhage region (i.e., the rest of the

TABLE I
NMRSE VALUES OF QSM RESULTS FOR A HEMORRHAGIC BRAIN.

|  | HEMORRHAGE AREA | REMAINING AREA | FULL BRAIN |
|---|---|---|---|
| U-net | 1.11 | 1.52 | 1.34 |
| AFTER-QSM | 0.38 | 0.37 | 0.38 |
| DIP | 0.32 | 0.54 | 0.45 |
| FINE | 0.32 | 0.59 | 0.49 |
| iLSQR | 0.25 | 0.45 | 0.37 |
| MEDI | 0.99 | 0.45 | 0.75 |
| **MoDIP** | **0.19** | **0.29** | **0.25** |

The hemorrhage area corresponds to the red box in Figure 6.

brain outside of the red box), and the entire brain.

Different methods were further compared on an in-vivo subject with cavernous hemangioma, scanned at 3T with an anisotropic spatial resolution ($\vec{v} = [0.88, 0.88, 2]$ mm) in a slightly titled acquisition orientation ($\vec{p} = [0.02, -0.12, 0.99]$). The local field map was reconstructed using the iQFM method [47] from the raw phase in a single step. Fig. 7 displays the reconstruction results in three orthogonal views, with red arrows pointing to noticeable artifacts. Similar to the simulated hemorrhagic case, MoDIP and AFTER-QSM produced the most visually appealing susceptibility maps, exhibiting reduced streaking artifacts near the hemangioma while maintaining excellent susceptibility contrast in healthy tissues. Moreover, MoDIP yielded the highest hemangioma susceptibility, consistent with the simulated hemorrhagic results in Fig. 6.

compared to all other network-based methods by a substantial margin. These results highlighted MoDIP's clear distinction from data-driven approaches and its ability to perform robustly across various test datasets.

### D. Overfitting and Stopping Criterion

As noted in the original DIP paper [44], excessive iterations

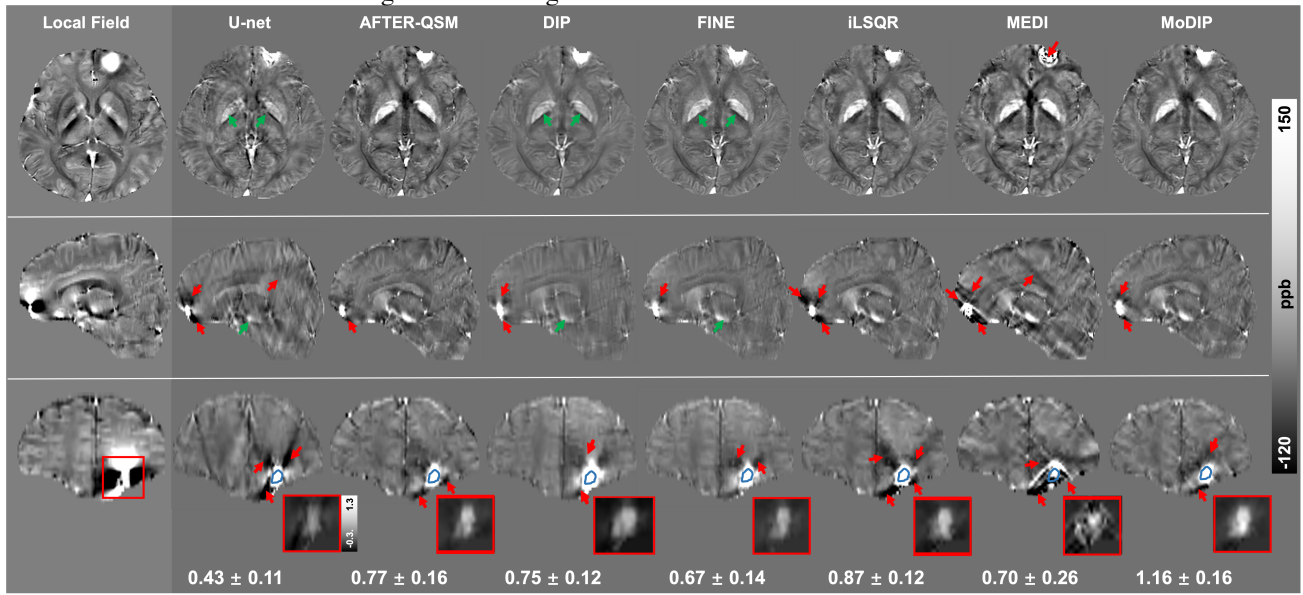

Figure. 7. QSM results of an in-vivo human brain ($\vec{v} = [0.875, 0.875, 2]$ mm, $\vec{p} = [0.02, -0.12, 0.99]$) with a cavernous hemangioma using different methods. Red arrows indicate apparent artifacts, while green arrows highlight the suppressed susceptibility tissue contrast. The hemangioma is delineated and measured using the blue contour, with a zoomed-in view and adjusted window level presented in the bottom row. The means and standard deviations of the hemangioma's susceptibility measurements are reported below the images.

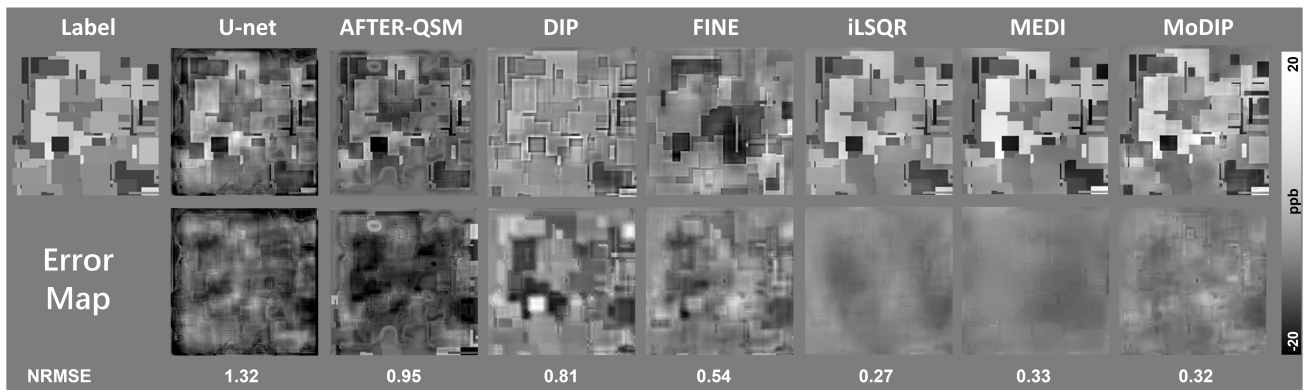

Figure. 8. QSM reconstruction results of a digital geometric phantom containing 80 cuboids using different methods. Error maps and the NRMSE values are shown below.

### C. Digital Geometric Phantom

Fig. 8 investigates the generalizability of different methods on a simulated geometric phantom. The phantom consists of 800 cuboids with side lengths randomly chosen from the range [1, 64] voxels, and each cuboid was assigned a uniform susceptibility value randomly selected from [-0.02, 0.02] ppm. These geometric susceptibility sources were arbitrarily placed and overlaid in an image of size 128×128×128, with susceptibility 0 set as the background.

MoDIP demonstrated comparable performance to conventional non-deep-learning methods (i.e., iLSQR and MEDI) that employ manually crafted regularizations designed to align with the sparsity characteristics of the simple geometric phantom. Additionally, MoDIP exhibited superior performance may lead to network overfitting and degraded image quality. Therefore, stopping the network weights optimization process in time is critical. To investigate the effects of network size on overfitting and determine the stopping criterion, we conducted a comparative study using an in-vivo brain dataset. This analysis involved comparing the outcomes of our proposed MoDIP method, which employed a mini-U-net architecture, with those of a modified variant using a U-net architecture with a pooling depth of 4. This modified variant is referred to as MoDIP-D4, resembling the U-net architecture utilized in DIP and FINE. We also performed Pure-DFO to illustrate the detrimental artifacts due to overfitting.

Fig. 9 shows the MoDIP, MoDIP-D4, and Pure-DFO results at iteration numbers 50, 100, 200, 400, 800, and 1000. It is evident that MoDIP-D4 in early iterations showed suppressed susceptiblity contrasts than MoDIP. However, as the number of iterations increased, MoDIP-D4 showcased more pronounced shadow artifacts, underscoring the importance of a suitable stopping mechanism. This observation aligns with findings from the original DIP paper. In contrast, MoDIP demonstrated faster convergence to an optimally regularized solution without introducing artifacts during the iterations. These results suggest that our proposed MoDIP approach with a mini-U-net demands less computational memory, converges more rapidly, and exhibits reduced possibility to overfitting, eliminating the need for early stopping. In practice, MoDIP inference can be terminated after a certain number of iterations or when the relative difference in total loss between two consecutive iterations falls below a predetermined threshold. Our testing across various cases indicates that 200 iterations strike a good balance between reconstruction accuracy and efficiency.

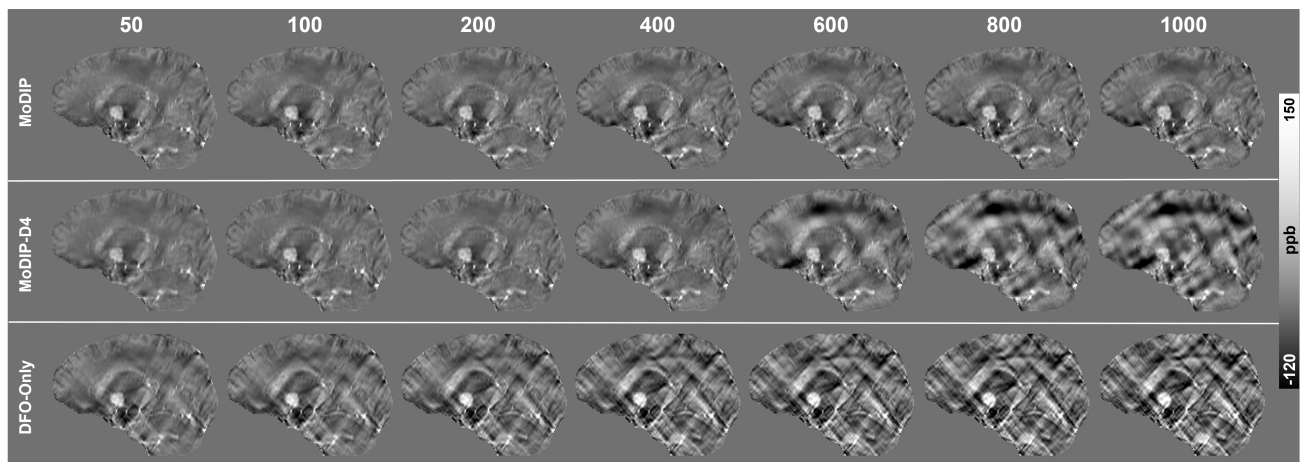

Figure. 9. Progression of reconstruction with increasing numbers of iterations on an in-vivo human brain with $\vec{v} = [1, 1, 1]$ mm and head orientation $\vec{p} = [0, 0, 1]$. Comparisons are presented for MoDIP, MoDIP-D4 (a modified variant employing a U-net architecture with a pooling depth of 4), and DFO-only methods.

## V. Discussion

Previous work [44] has shown that the performance of DIP, partially determined by the extent of the implicit prior regularization, is highly influenced by the selection of network architectures and the number of iterations. Solely optimizing the network weights against the model loss faces limitations in reaching the optimal solution or risk overfitting with excessive iterations, undermining the benefits of DIP. These deep and large networks also require extensive GPU memory and converge slowly, making them impractical for dealing with full-sized 3D high-resolution volumes in QSM dipole inversion.

MoDIP elevates the DIP concept by coupling it with conventional optimization to address these issues. Instead of relying on deep and large networks, we employ a shallow and small network and guide it towards the QSM solutions by optimizing directly on the image, without involving network weights. The theory behind MoDIP is that a small network can more easily and quickly converge to a suitable intermediate state for subsequent physical model-based optimization, compared to a deep network directly converging to the QSM solution. In MoDIP, both neural network updates and DFO steps are critical in reaching the optimal solution, sharing the workload, and reducing task complexity. The network focuses on prior regularization, while image optimization enhances data fidelity. This idea aligns with unrolled deep learning methods like LPCNN [40] and MoDL-QSM [39]. However, in those methods, the network regularizers are learned from the training datasets and are explicitly applied to regularize the QSM images, which makes them less generalizable than MoDIP to various testing datasets. In theory, it is possible to fine-tune the pre-trained unrolled models during inference. However, this approach would be highly computationally demanding due to the involvement of multiple large networks. Similar to FINE, fine-tuning unrolled models becomes challenging when test sets substantially differ from the training sets.

Ablation studies revealed that the small network not only made MoDIP more practical and efficient with significantly less GPU memory requirement but also helped alleviate the risk of overfitting over iterations. Replacing random noise as network input with the local field measurement contributes significantly to MoDIP's performance, possibly because the network benefits from shared image features between the local field input and the susceptibility map output. Pre-training the network weights with paired local field and QSM datasets (FINE) showed limited improvement over untrained DIP networks and could even reduce its generalizability to different testing sets.

We also compared MoDIP with conventional non-deep-learning methods, as well as supervised deep-learning methods. These evaluation datasets included healthy and pathological brains and a digital geometric phantom, varying in object shape, susceptibility range, and acquisition parameters. MoDIP achieved outstanding generalizability in accurately reconstructing susceptibility maps from these diverse testing datasets. Notably, MoDIP achieved these results in under 4.5 minutes without the need for pre-training on paired datasets, exhibiting its practicality and flexibility in real-world applications.

Our work has identified several limitations of MoDIP as well. For instance, MoDIP does not account for local field pre-processing steps, such as brain extraction, phase unwrapping, and background field removal. As a result, the method may be susceptible to accumulated errors from these prior steps. Future work could explore a similar approach to implement MoDIP based on the physical model between the raw phase and QSM [50]. Another limitation is that the computational cost of MoDIP is still higher than conventional iterative methods. Specifically, MoDIP takes nearly twice as long as iLSQR for reconstruction, using the mini-U-net starting from 32 channels. One possible solution to mitigate this computational burden is to reduce the number of network channels while increasing the number of DFO steps, although this may result in a trade-off with performance. Future work could find ways to balance computational efficiency and performance in MoDIP.

## VI. CONCLUSION

In this work, we propose MoDIP, an unsupervised deep learning method based on the DIP concept, specifically designed to address the challenging task of 3D high-resolution QSM dipole inversion. MoDIP represents an advancement in this field, harnessing the benefits of a compact untrained network and a data fidelity optimization process. Consequently, the method reduces computational costs and demonstrates exceptional generalizability and robustness in accurately reconstructing susceptibility maps from various challenging datasets without requiring any training process. MoDIP holds great promise for practical applications in clinical settings.

## REFERENCE

[1] X. Li, D. S. Vikram, I. A. L. Lim, C. K. Jones, J. A. Farrell, and P. C. van Zijl, "Mapping magnetic susceptibility anisotropies of white matter in vivo in the human brain at 7 T," *Neuroimage,* vol. 62, no. 1, pp. 314-330, 2012.
[2] H. Sun *et al.*, "Validation of quantitative susceptibility mapping with Perls' iron staining for subcortical gray matter," *Neuroimage,* vol. 105, pp. 486-492, 2015.
[3] F. Schweser, A. Deistung, B. W. Lehr, and J. R. Reichenbach, "Differentiation between diamagnetic and paramagnetic cerebral lesions based on magnetic susceptibility mapping," *Medical physics,* vol. 37, no. 10, pp. 5165-5178, 2010.
[4] J. Acosta-Cabronero, G. B. Williams, A. Cardenas-Blanco, R. J. Arnold, V. Lupson, and P. J. Nestor, "In vivo quantitative susceptibility mapping (QSM) in Alzheimer's disease," *PLoS One,* vol. 8, no. 11, p. e81093, 2013, doi: 10.1371/journal.pone.0081093.
[5] J. Acosta-Cabronero *et al.*, "The whole-brain pattern of magnetic susceptibility perturbations in Parkinson's disease," *Brain,* vol. 140, no. 1, pp. 118-131, Jan 2017, doi: 10.1093/brain/aww278.
[6] J. M. van Bergen *et al.*, "Quantitative Susceptibility Mapping Suggests Altered Brain Iron in Premanifest Huntington Disease," *AJNR Am J Neuroradiol,* vol. 37, no. 5, pp. 789-96, May 2016, doi: 10.3174/ajnr.A4617.
[7] A. M. Elkady, D. Cobzas, H. Sun, P. Seres, G. Blevins, and A. H. Wilman, "Five year iron changes in relapsing-remitting multiple sclerosis deep gray matter compared to healthy controls," *Mult Scler Relat Disord,* vol. 33, pp. 107-115, Aug 2019, doi: 10.1016/j.msard.2019.05.028.
[8] A. M. Elkady, D. Cobzas, H. Sun, G. Blevins, and A. H. Wilman, "Discriminative analysis of regional evolution of iron and myelin/calcium in deep gray matter of multiple sclerosis and healthy subjects," *J Magn Reson Imaging,* Mar 14 2018, doi: 10.1002/jmri.26004.
[9] T. Liu, K. Surapaneni, M. Lou, L. Cheng, P. Spincemaille, and Y. Wang, "Cerebral microbleeds: burden assessment by using quantitative susceptibility mapping," *Radiology,* vol. 262, no. 1, pp. 269-278, 2012.
[10] W. Chen *et al.*, "Intracranial calcifications and hemorrhages: characterization with quantitative susceptibility mapping," *Radiology,* vol. 270, no. 2, pp. 496-505, Feb 2014, doi: 10.1148/radiol.13122640.
[11] H. Sun *et al.*, "Quantitative Susceptibility Mapping for Following Intracranial Hemorrhage," *Radiology,* vol. 288, no. 3, pp. 830-839, Sep 2018, doi: 10.1148/radiol.2018171918.
[12] A. De, H. Sun, D. J. Emery, K. S. Butcher, and A. H. Wilman, "Rapid quantitative susceptibility mapping of intracerebral hemorrhage," *Journal of Magnetic Resonance Imaging,* vol. 51, no. 3, pp. 712-718, 2020.
[13] D. Z. Balla *et al.*, "Functional quantitative susceptibility mapping (fQSM)," *Neuroimage,* vol. 100, pp. 112-124, 2014.
[14] J. Zhang, T. Liu, A. Gupta, P. Spincemaille, T. D. Nguyen, and Y. Wang, "Quantitative mapping of cerebral metabolic rate of oxygen (CMRO2) using quantitative susceptibility mapping (QSM)," *Magnetic resonance in medicine,* vol. 74, no. 4, pp. 945-952, 2015.
[15] Y. Ma, H. Sun, J. Cho, E. L. Mazerolle, Y. Wang, and G. B. Pike, "Cerebral OEF quantification: A comparison study between quantitative susceptibility mapping and dual‐gas calibrated BOLD imaging," *Magnetic resonance in medicine,* vol. 83, no. 1, pp. 68-82, 2020.
[16] Y. Ma, E. L. Mazerolle, J. Cho, H. Sun, Y. Wang, and G. B. Pike, "Quantification of brain oxygen extraction fraction using QSM and a hyperoxic challenge," *Magnetic resonance in medicine,* vol. 84, no. 6, pp. 3271-3285, 2020.
[17] H. Sun, P. Seres, and A. Wilman, "Structural and functional quantitative susceptibility mapping from standard fMRI studies," *NMR in Biomedicine,* vol. 30, no. 4, p. e3619, 2017.
[18] K. Shmueli, J. A. de Zwart, P. van Gelderen, T. Q. Li, S. J. Dodd, and J. H. Duyn, "Magnetic susceptibility mapping of brain tissue in vivo using MRI phase data," *Magnetic Resonance in Medicine: An Official Journal of the International Society for Magnetic Resonance in Medicine,* vol. 62, no. 6, pp. 1510-1522, 2009.
[19] Y. Wang and T. Liu, "Quantitative susceptibility mapping (QSM): decoding MRI data for a tissue magnetic biomarker," *Magnetic resonance in medicine,* vol. 73, no. 1, pp. 82-101, 2015.
[20] F. Schweser, A. Deistung, B. W. Lehr, and J. R. Reichenbach, "Quantitative imaging of intrinsic magnetic tissue properties using MRI signal phase: an approach to in vivo brain iron metabolism?," *Neuroimage,* vol. 54, no. 4, pp. 2789-2807, 2011.
[21] B. Bilgic *et al.*, "Fast quantitative susceptibility mapping with L1-regularization and automatic parameter selection," *Magn Reson Med,* vol. 72, no. 5, pp. 1444-59, Nov 2014, doi: 10.1002/mrm.25029.
[22] F. Schweser, K. Sommer, A. Deistung, and J. R. Reichenbach, "Quantitative susceptibility mapping for investigating subtle susceptibility variations in the human brain," *Neuroimage,* vol. 62, no. 3, pp. 2083-100, Sep 2012, doi: 10.1016/j.neuroimage.2012.05.067.
[23] C. Kames, V. Wiggermann, and A. Rauscher, "Rapid two-step dipole inversion for susceptibility mapping with sparsity priors," *Neuroimage,* vol. 167, pp. 276-283, Feb 15 2018, doi: 10.1016/j.neuroimage.2017.11.018.
[24] D. Khabipova, Y. Wiaux, R. Gruetter, and J. P. Marques, "A modulated closed form solution for quantitative susceptibility mapping—a thorough evaluation and comparison to iterative methods based on edge prior knowledge," *NeuroImage,* vol. 107, pp. 163-174, 2015.
[25] T. Liu *et al.*, "Morphology enabled dipole inversion (MEDI) from a single-angle acquisition: comparison with COSMOS in human brain imaging," *Magn Reson Med,* vol. 66, no. 3, pp. 777-83, Sep 2011, doi: 10.1002/mrm.22816.
[26] L. Bao, X. Li, C. Cai, Z. Chen, and P. C. van Zijl, "Quantitative Susceptibility Mapping Using Structural Feature Based Collaborative Reconstruction (SFCR) in the Human Brain," *IEEE Trans Med Imaging,* vol. 35, no. 9, pp. 2040-50, Sep 2016, doi: 10.1109/TMI.2016.2544958.
[27] W. Li *et al.*, "A method for estimating and removing streaking artifacts in quantitative susceptibility mapping," *Neuroimage,* vol. 108, pp. 111-22, Mar 2015, doi: 10.1016/j.neuroimage.2014.12.043.
[28] H. Wei *et al.*, "Streaking artifact reduction for quantitative susceptibility mapping of sources with large dynamic range," *NMR in Biomedicine,* vol. 28, no. 10, pp. 1294-1303, 2015.
[29] H. Sun, Y. Ma, M. E. MacDonald, and G. B. Pike, "Whole head quantitative susceptibility mapping using a least-norm direct dipole inversion method," *Neuroimage,* vol. 179, pp. 166-175, 2018.
[30] Z. Liu, Y. Kee, D. Zhou, Y. Wang, and P. Spincemaille, "Preconditioned total field inversion (TFI) method for quantitative susceptibility mapping," *Magn Reson Med,* vol. 78, no. 1, pp. 303-315, Jul 2017, doi: 10.1002/mrm.26331.

[31] C. Langkammer *et al.*, "Fast quantitative susceptibility mapping using 3D EPI and total generalized variation," *Neuroimage,* vol. 111, pp. 622-630, 2015.

[32] Y. Chen, A. Jakary, S. Avadiappan, C. P. Hess, and J. M. Lupo, "QSMGAN: Improved Quantitative Susceptibility Mapping using 3D Generative Adversarial Networks with increased receptive field," *Neuroimage,* vol. 207, p. 116389, Feb 15 2020, doi: 10.1016/j.neuroimage.2019.116389.

[33] Y. Gao *et al.*, "xQSM: quantitative susceptibility mapping with octave convolutional and noise-regularized neural networks," *NMR Biomed,* vol. 34, no. 3, p. e4461, Mar 2021, doi: 10.1002/nbm.4461.

[34] J. Yoon *et al.*, "Quantitative susceptibility mapping using deep neural network: QSMnet," *Neuroimage,* vol. 179, pp. 199-206, Oct 1 2018, doi: 10.1016/j.neuroimage.2018.06.030.

[35] W. Jung *et al.*, "Exploring linearity of deep neural network trained QSM: QSMnet+," *Neuroimage,* vol. 211, p. 116619, 2020.

[36] J. Liu and K. M. Koch, "Meta-QSM: An image-resolution-arbitrary network for QSM reconstruction," *arXiv preprint arXiv:1908.00206,* 2019.

[37] Z. Li *et al.*, "Meta-learning based interactively connected clique U-net for quantitative susceptibility mapping," *IEEE Transactions on Computational Imaging,* vol. 7, pp. 1385-1399, 2021.

[38] S. Bollmann *et al.*, "DeepQSM - using deep learning to solve the dipole inversion for quantitative susceptibility mapping," *Neuroimage,* vol. 195, pp. 373-383, Jul 15 2019, doi: 10.1016/j.neuroimage.2019.03.060.

[39] R. Feng *et al.*, "MoDL-QSM: Model-based deep learning for quantitative susceptibility mapping," *Neuroimage,* vol. 240, p. 118376, Oct 15 2021, doi: 10.1016/j.neuroimage.2021.118376.

[40] K. W. Lai, M. Aggarwal, P. van Zijl, X. Li, and J. Sulam, "Learned Proximal Networks for Quantitative Susceptibility Mapping," *Med Image Comput Comput Assist Interv,* vol. 12262, pp. 125-135, Oct 2020, doi: 10.1007/978-3-030-59713-9_13.

[41] D. Polak *et al.*, "Nonlinear dipole inversion (NDI) enables robust quantitative susceptibility mapping (QSM)," *NMR Biomed,* vol. 33, no. 12, p. e4271, Dec 2020, doi: 10.1002/nbm.4271.

[42] Z. Xiong, Y. Gao, F. Liu, and H. Sun, "Affine transformation edited and refined deep neural network for quantitative susceptibility mapping," *NeuroImage,* vol. 267, p. 119842, 2023.

[43] G. Oh, H. Bae, H. S. Ahn, S. H. Park, W. J. Moon, and J. C. Ye, "Unsupervised resolution-agnostic quantitative susceptibility mapping using adaptive instance normalization," *Med Image Anal,* vol. 79, p. 102477, Jul 2022, doi: 10.1016/j.media.2022.102477.

[44] D. Ulyanov, A. Vedaldi, and V. Lempitsky, "Deep image prior," in *Proceedings of the IEEE conference on computer vision and pattern recognition*, 2018, pp. 9446-9454.

[45] J. Zhang *et al.*, "Fidelity imposed network edit (FINE) for solving ill-posed image reconstruction," *Neuroimage,* vol. 211, p. 116579, May 1 2020, doi: 10.1016/j.neuroimage.2020.116579.

[46] T. Liu, P. Spincemaille, L. de Rochefort, B. Kressler, and Y. Wang, "Calculation of susceptibility through multiple orientation sampling (COSMOS): a method for conditioning the inverse problem from measured magnetic field map to susceptibility source image in MRI," *Magn Reson Med,* vol. 61, no. 1, pp. 196-204, Jan 2009, doi: 10.1002/mrm.21828.

[47] Y. Gao *et al.*, "Instant tissue field and magnetic susceptibility mapping from MRI raw phase using Laplacian enhanced deep neural networks," *Neuroimage,* vol. 259, p. 119410, Oct 1 2022, doi: 10.1016/j.neuroimage.2022.119410.

[48] H. S. Abdul-Rahman, M. A. Gdeisat, D. R. Burton, M. J. Lalor, F. Lilley, and C. J. Moore, "Fast and robust three-dimensional best path phase unwrapping algorithm," *Applied optics,* vol. 46, no. 26, pp. 6623-6635, 2007.

[49] H. Sun and A. H. Wilman, "Background field removal using spherical mean value filtering and Tikhonov regularization," *Magnetic resonance in medicine,* vol. 71, no. 3, pp. 1151-1157, 2014.

[50] T. Liu, D. Zhou, P. Spincemaille, and Y. Wang, "Differential approach to quantitative susceptibility mapping without background field removal," in *Proceedings of the 22nd Annual Meeting of ISMRM, Milan, Italy*, 2014, vol. 597.